# Rapid Lung MRI @3T in ILD Patients: A Feasibility Study

Bingjie Yang[1,2,5], Martina Büttner[3], Hanna Frantz[2], Patrick Metze[2], Viktoria Haiss[2], Gerlinde Schmidtke-Schrezenmeier[4], Cornelia Kropf-Sanchen[4], Meinrad Beer[3], Wolfgang Rottbauer[2], Volker Rasche[2]

[1]Department of Ultrasound, Zhong Da Hospital, Medical School, Southeast University, 87 Dingjiaqiao Road, Nanjing, Jiangsu Province, 210009, China
[2]Department of Internal Medicine II, Experimental Cardiovascular Imaging, Ulm University Medical Centre, Ulm, Germany;
[3]Department of Radiology, Ulm University Medical Centre, Ulm, Germany
[4]Department of Internal Medicine II, Pulmonology, Ulm University Medical Centre, Ulm, Germany;
[5]Medical School, Southeast University, 87 Dingjiaqiao Road, Nanjing, Jiangsu Province, 210009, China

**Abstract**

**Background:** Computed tomography (CT) remains the gold standard imaging method for interstitial lung disease (ILD) evaluation. However, the main disadvantage of CT is the potential radiation exposure, particularly considering the chronicity of ILD and the requirement for frequent follow-up examinations. The application of the ultra-fast steady-state free precession (ufSSFP) pulse sequence for FD MRI at 3T appears problematic because of the specific absorption rate (SAR), which reduces the maximal achievable flip angle, as well as increased susceptibility effects, resulting in more severe banding. This study aimed to assess the diagnostic utility of a conventional FLASH technique at 3T MRI in the detection of ILD patients in combination with functional and morphological information by a rather simple but straightforward approach for SNR improvement by simple averaging and to compare it with the current imaging gold standard, CT.

**Methods:** Coronal FLASH MR images with a temporal resolution of 272 ms of healthy volunteers and patients with ILD were acquired at 3T. In patients additional CT images were acquired during clinical routine. Average expiration or inspiration images were derived from the continuous data acquisition. Proton fraction, signal-to-noise ratio, and fractional ventilation were calculated. Concordance and agreement between CT and MRI were assessed.

**Results:** The analysis of morphological abnormalities of the lung parenchyma in the ILD patients revealed  low for traction bronchiectasis (kappa = 0.357) to excellent for pleural plaque (kappa = 0.857) agreement. Proton fraction of posterior slices of the lung showed a significant difference between healthy volunteers and ILD patients (p < 0.01). Fractional ventilation maps are homogeneous in healthy volunteers while it shows regions with reduced fractional ventilation in patients with interstitial lung diseases, which are matched with areas with morphological changes.

**Conclusions:** This study shows the feasibility of a applying a rather simple real-time MRI protocol with subsequent averaging to obtain morphological and regional ventilation information in patients with interstitial lung diseases, which may contribute to further clinical translation of lung MRI.

**Keywords:** real-time MRI; interstitial lung disease; proton fraction; fractional ventilation

**1. Introduction**

Interstitial lung diseases (ILD) represent a heterogeneous group of disorders with diffuse lung involvement, mainly affecting the most peripheral and delicate interstitium in the alveolar walls (1-3). It concerns alveolar epithelium, pulmonary capillary endothelium, basement membrane, and perivascular and perilymphatic tissues. The disorders frequently affect not only the interstitium, but also the airspaces, peripheral airways, and vessels along with their respective epithelial and endothelial linings. There are more than 200 ILDs, including immunologic, infectious, environmental, toxic, and genetic mechanisms, whereas an etiological agent cannot be identified in many ILDs (4-7).

Most patients with suspected ILD are likely to undergo pulmonary function testing (PFT), which is useful in the diagnosis and assessment of the severity of the disease (8-10). However, PFT is effort dependent and requires patient cooperation and understanding in performing the tests for accurate results. Further it is limited to provide global information on lung function only, which especially in cases with local compensation by healthy neighboring parenchyma is critical. Tissue biopsy provides a definitive diagnosis but has the disadvantage of being invasive (11). Considering the limitations of all these currently available investigations, imaging plays an important role in the diagnosis of ILD.

Even though only providing information on lung morphology, computed tomography (CT) remains the gold standard imaging method for ILDs evaluation as clinical presentation and histopathologic patterns show significant concordance in ILDs (12). Its main disadvantage is the potential radiation exposure, particularly considering the chronicity of ILD and the requirement for frequent follow-up examinations (13,14).

In recent years, magnetic resonance imaging (MRI) has turned out as an attractive radiation-free alternative to CT. MRI with inhaled hyperpolarized noble gases ($^3$He, $^{129}$Xe) was introduced for the assessment of static and dynamic lung ventilation (15,16), even enabling quantification of gas transfer ($^{129}$Xe) (17). However, due to the complexity of the imaging setup and potential impact of the different atomic weight of the agents, there is no broad clinical use of the technique. Alternatively, proton MRI using a T1-shortening contrast agent, such as oxygen for ventilation, or gadolinium-based contrast agents for perfusion was used for functional assessment (18-20). Even though promising, the technique has not yet reached clinical routine, due to the demand of well-controlled oxygen ventilation in the scanner and the possible risk of acute allergic reactions and nephrogenic systemic fibrosis in patients with impaired renal function (21), especially in case of longitudinal disease monitoring.

Further, attractive MR methods have been introduced using the intrinsic signal intensity changes of the parenchyma over the respiratory (changes in air volume) and circulatory (changes in amount of unsaturated blood) cycle. These approaches either use realtime data acquired continuously over multiple respiratory cycles (22,23), self-gated data acquisitions (24), or breathhold (BH) acquisitions in inspiration and expiration (25,26). For data analysis either direct comparison of the relative intensity changes (25) or spectral analysis of the intensity changes over the respiratory and cardiac cycle (FD, SENSEFUL, PREFUL) (24,27-29) have been suggested. One major limitation in lung imaging results from the very low tissue density (about 0.1 g/cm$^3$) (30) and ultrashort T2* relaxation times due to the multiple air-tissue interfaces. Bauman et al. addressed this limitations at 1.5T by optimization of an ultra-fast steady-state free precession (ufSSFP) scheme in combination with a Fourier decomposition (FD) analysis (27). The optimized ufSSFP sequence benefits from the short TR values and improves the image quality by increasing the signal intensity in the lung parenchyma while reducing the amount of banding artifacts at 1.5T. While in general the signal should increase at 3T, translation of the ufSSFP pulse sequence to higher field strength appears problematic. Major limitations arise from specific absorption rate (SAR), which reduces the maximal achievable flip angle, and more importantly increased susceptibility effects, resulting in severe banding artifacts. Even though lung images with ufSSFP have been reported a markedly inferior quality as compared to 1.5Twas shown (27).

The objective of this work is to investigate a straightforward approach for SNR improvement in free-breathing (FB) real-time lung imaging at 3T based on a conventional FLASH sequence and to initially assess its diagnostic potential in ILD patients in direct comparison to HRCT.

## 2. Materials and methods

*2.1 Subjects*

The study was conducted in accordance with the Declaration of Helsinki, and the protocol was approved by the institutional ethics board of Ulm University (number:185/19 ). Written informed consent was obtained from all subjects prior to enrolment in the study. Nine healthy volunteers (mean age: 27.7 ± 1.5), 5 males and 4 females, with no reported history of cardiopulmonary disease or any pulmonary event (such as infection) within one month prior to the examination were included. Sixteen patients diagnosed with ILD (mean age: 62.3 ± 13.6), 10 males and 6 females, were referred for idiopathic pulmonary fibrosis (n = 3), sarcoidosis (n = 5), scleroderma (n = 3), cryptogenic organizing pneumonia (n = 1), autoimmune diseases (n = 1), Sharp-Syndrome (n = 1) and hypersensitivity pneumonitis (n = 2). All ILD diagnoses of patients were made by consensus of a board of certified pulmonologists, radiologists, and pathologists based on the results of pulmonary function and laboratory tests, clinical findings, thin-section CT, and pathological examination findings. Exclusion criteria were any features of malignancy involving the lungs and any contraindication for MRI imaging.

*2.2 Image Acquisition*

*CT examination*

All patients underwent a non-contrast enhanced CT chest scan during routine clinical workup. Data were acquired with a 64-detector row CT with acquisition parameters as: 38 mm × 0.6 mm collimation, pitch 1.7, 512 × 512 matrix, 120 kV, 270 mA, and 1 mm reconstructed slice thickness. The scan was performed during full inspiration covering the whole lung from apex to the diaphragm.

*Real-time MRI*

All MRI data were acquired on a 3T whole-body system (Achieva 3T, Philips Healthcare, Best, The Netherlands) with the integrated 16-channel (4×4) posterior coil in combination with a matching

16-channel (4×4) anterior segment. Data were acquired in three coronal slices, with the midslice cantered on the tracheal bifurcation. A spoiled gradient echo sequence (FLASH) was used with the following settings: field of view (FOV) 400 × 400 mm$^2$, matrix size 144 × 144, slice thickness 15 mm, echo time (TE) 0.60 msec, repetition time (TR) 1.89 msec, flip angle 4°, bandwidth 357 Hz/px, temporal resolution 272 msec. For each slice, data was continuously acquired for about 1 minute, yielding roughly 200 images. The overall examination time resulted into less than 10 minutes including planning.

*2.3 Postprocessing*

MRI images were reconstructed with Gyrotools (Gyrotools LLC, Zurich, Switzerland) and post-processed with an in-house developed framework implemented in Matlab (MathWorks, Natick, Massachusetts, USA). The first eight images of each time series were discarded to avoid inclusion of images acquired during the transient state of steady-state build-up. For SNR improvement, prior to the subsequent analysis images acquired in the same respiratory stage were identified and averaged applying an image-based navigator-like signal which was derived from a manually identified region of interest (ROI) covering the liver-lung interface (Figure 1).

*2.4 Image analysis*

Analysis was performed blinded to the clinical data in random order by a radiologist with 6 years of experience in thoracic imaging to evaluate the anatomical information, the presence of the structural abnormalities of the lung was scored on a yes/no scale for each slice of every patient and comprised honeycombing, septal lines, reticulation, consolidation, ground-glass opacity, nodules, cysts, traction bronchiectasis, pleural effusion and emphysema as defined by the glossary of terms from Fleishner Society (31).

MRI data analysis was performed on magnitude data. The mean noise $\overline{SI}_{noise}$ was derived from a manually identified artifact-free background and subtracted prior to quantitative analysis to avoid a substantial impact of the noise floor to the low-apparent SNR data analysis (28). Lung parenchyma was segmented semi-automatically to carefully exclude larger vessels. Segmentation was performed independently in all images by an active contour method, based on the Chan-Vese technique (32) (Image Segmenter App, MatLab). Lung parenchyma voxels apparent SNR, proton fraction ($f_P$), and FV were quantified (25).

In brief, the apparent SNR in the lung parenchyma was calculated as:

$$SNR = \frac{\overline{SI}_{lung}}{\sigma_{noise}}$$

with $\overline{SI}_{lung}$ being the mean value of lung parenchyma and $\sigma_{noise}$ the standard deviation of the previously identified background region. $F_P$ was calculated pixel-wise as:

$$f_p = \frac{\overline{SI}_{lung}}{\overline{SI}_{muscle}} \cdot exp\left(\frac{TE}{T_2^*}\right)$$

with $\overline{SI}_{muscle}$ being the intramuscular signal reference derived from a manually identified ROI placed in the intercostal muscle, and $T_2^*$ = 0.74 used for correcting $T_2^*$ effects caused by the rapid signal decay at 3T (33). The FV (34) map was derived from the relative change between IN and EX signal intensities as:

$$FV = \frac{\overline{SI}_{EX} - \overline{SI}_{IN}}{\overline{SI}_{EX}}$$

To ensure a proper match of the lung parenchyma between IN and EX, a nonrigid image registration was performed by applying the MIRT—Medical Image Registration Toolbox for MatLab (35) with the sum of squared differences (SSD) as a similarity measure in a free-form deformation algorithm. All images were registered to the expiration image.

*2.5 Statistical analysis*

All reconstructed images were normalized before the analysis and all continuous data are shown as mean ± standard deviation (SD). Statistical significances between the different groups were assessed by using an analysis of variance (ANOVA) test. A P value below 0.05 was considered to be statistically significant. Agreement between CT and MRI, was assessed by using the kappa test for categoric variable. Kappa values were classified as null (0), slight (0–0.20), fair (0.21–0.40), moderate (0.41–0.60), good (0.61–0.80), and very good (0.81–1) (36). The anatomical information of MRI was assessed twice, prior (blinded) and after (unblinded) CT reading. Compared to CT, the overall sensitivity, specificity, PPV and NPV of MRI were calculated. Data were analyzed with GraphPad Prism (LLC, BOSTON, USA) and SPSS (IBM Corporation, Armonk, NY, USA).

**3. Results**

The MRI protocol (volunteers, n=9; patients, n=16) and HRCT (patients, n=16) was completed in all enrolled volunteers and patients. The real-time image quality was sufficient for image-based self-gating and a high-quality respiratory navigator signal could be derived for SNR improvement of images obtained in different respiratory phases (Figure 2). A mean SNR improvement in the lung parenchyma from 12.8+/-6.8 to 21.8+/-14.1 could be achieved by the proposed approach as exemplary shown in Figure 3 in one ILD patient with idiopathic pulmonary fibrosis.

In direct comparison to HRCT, interstitial-related morphological alteration of the lung (Figure 4) could be assessed on the averaged real-time MRI with low agreement for honeycombing and traction bronchiectasis (kappa = 0.333) to excellent agreement for nodule, cyst and pleural effusion (kappa = 1) (Table 1) for blinded reading of MRI and moderate agreement for ground glass opacity (kappa = 0.458) to excellent agreement for nodule, cyst and pleural effusion (kappa = 1) (Table 2) for unblinded reading of MRI. The overall sensitivity, specificity, PPV and NPV were 0.63, 0.99, 0.96, 0.87 for blinded and 0.74, 0.99, 0.97, 0.90 for unblinded reading of MRI.

Differences in the lung parenchyma normalized signal intensities between expiration and inspiration were observed in all slices of all volunteers and patients. In all subjects the proton fraction in expiration resulted significantly ($p < 0.001$) higher than in inspiration and increased from anterior to posterior (Figure 5). There was a general trend to increased proton densities in the ILD patient group with significant ($p < 0.01$) differences for the posterior slice. Further the interquartile range was higher in the ILD group (Figure 5). The mean value of proton fraction over all coronal slices was $0.21 \pm 0.05$ for healthy volunteers and $0.32 \pm 0.13$ for ILD patients. A significant reduction of the SNR from anterior to posterior was observed in all subjects. The SNR and its interquartile range were higher in ILD patients compared with healthy volunteers (Figure 6).

Fractional ventilation (FV) maps could successfully be obtained from the analysis of the signal intensity changes between inspiration and expiration in all volunteers and patients. Real-time MRI on coronal planes shows mostly homogeneous FV within lungs in healthy volunteers (Figure 7) while heterogeneous FV was apparent in ILD patients with regions of reduced FV matching areas with morphological. E.g. clear reduction of FV can be detected in the area of fibrotic changes, ground glass opacities and pleural plaque (Figure 8).

**4. Discussion**

This study shows that even at 3T lung MRI appears feasible with a rather simple straightforward low-resolution 2D FLASH protocol. Due to the rapid data acquisition in less than 10minutes on average and no required breathholds, the MRI examination was well tolerated by all volunteers and patients. Even though compromising spatial resolution, the combination of real-time imaging with temporal averaging demonstrated good imaging quality in interstitial lung disease with clinical information comparable to that of CT. Even though an excellent specificity, PPV and NPV was observed, the results still indicate a clear limitation of MRI regarding sensitivity. The possibility of additionally deriving functional parameter and the lack of ionizing radiation may make 3T MRI an attractive imaging alternative to HRCT especially for longitudinal monitoring of disease progression.

It Is feasible to obtain quantitative parameters of lung density and function applying a conventional spoiled gradient echo (FLASH) sequence in a combination with image-based motion-compensated averaging. Proton density (PD) quantification revealed increasing PD values from anterior to posterior. This has been reported in previous studies (25) and can most-likely be attributed to water redistribution in the lungs. More importantly a clear trend to increased PD was observed in the ILD patients, which may be caused by morphological changes, such as fibrosis, reticulations, and ground glass opacities as also clearly visible in the images. Some structural abnormalities could only be identified after reading of the respective CT data, which may be attributed to insufficient experience in reading lung MRI data. A general overestimation of the mean PD values in healthy volunteers of ~21% may be explained either by incorrect $T2^*$ values or by contributions of larger blood vessels due to the rather poor spatial resolution and especially

As shown in a previously reported study, tidal volume variability is a critical factor for fractional ventilation reproducibility (28,29), and to that regard free breathing approaches appear superior to breathhold techniques. Previously reported approaches either require complex self-gating techniques (24,28) for providing the required multi-respiratory phase data, or directly analyze the real-time image stream (22,27). Reliable segmentation of the lung and registration between the different respiratory stages is required to avoid falsified results by partial volume effects or blood vessels and comparison of nonmatching lung parenchyma (23,24). In both approaches prior to the analysis dozens or even hundreds of images have to be proper registered prior to performing the Fourier analysis. Even though the FD approach intrinsically improves SNR, the approach as such maybe error prone and may require time consuming corrections. In the suggested approach the required gain in SNR is achieved by averaging in the spatial domain and the relative fractional ventilation (FV) information derived from a single inspiration and expiration image, which substantially eases the registration task. In our study, during normal tidal breathing, the regions with reduced fractional ventilation in patients with interstitial lung diseases correlate well with those showing predominant disease patterns on CT and MRI. Regions with slightly increased fractional ventilation maybe attributed to compensation of regional function for the more diseased areas.

The study has several limitations. First, the applied three-slice 2D technique suffers from the intrinsic low spatial resolution especially into slice selection directions and the incomplete coverage of the entire lung. Even though this enables rapid free-breathing acquisition thus allowing scanning of patients with limited tolerance for lying flat in the scanner or not able to keep their breath, the sensitivity for slight morphological alterations needs to be further investigated.

The calculated proton density is dependent on proper identification of a muscle reference signal and lung parenchymal T2* value. Even though the T2* values has been derived for 3T in healthy volunteers (33), changes of the lung parenchyma density and B0 inhomogeneities impact the local T2* value and may cause falsified PD results. Further the required muscle reference value may be impacted by coil-sensitivity patterns and B1 inhomogeneities, the compensation of which requires accurate coil sensitivity information. However, due to the low intrinsic signal in the lungs deriving accurate coil sensitivity maps is challenging, which is further complicated by the large respiratory amplitudes as present in patients suffering from dyspnea. The related displacement of the anterior coil may demand respiration phase-dependent coilmaps for accurate PD estimation.

Further limitations include time-consuming still semiautomatic approaches for segmentation and registration and a rather low number of patients.

In conclusion, despite all the afore mentioned limitations, real-time MRI in combination with simple image-based averaging appears to be a promising approach for the evaluation of lung morphological and ventilation. It may contribute to further translation of lung MRI into the clinics for routine and research investigations of interstitial lung diseases.

**References**

1. Annesi-Maesano I, Lundbäck B, Viegi G. Respiratory Epidemiology: ERS Monograph. European Respiratory Society; 2014.

2. Nici L, Donner C, Wouters E, Zuwallack R, Ambrosino N, Bourbeau J, Carone M, Celli B, Engelen M, Fahy B. American thoracic society/European respiratory society statement on pulmonary rehabilitation. American journal of respiratory and critical care medicine 2006;173:1390-413.

3. Travis WD, Costabel U, Hansell DM, King Jr TE, Lynch DA, Nicholson AG, Ryerson CJ, Ryu JH, Selman M, Wells AU. An official American Thoracic Society/European Respiratory Society statement: update of the international multidisciplinary classification of the idiopathic interstitial pneumonias. American journal of respiratory and critical care medicine 2013;188:733-48.

4. Collard HR, Tino G, Noble PW, Shreve MA, Michaels M, Carlson B, Schwarz MI. Patient experiences with pulmonary fibrosis. Respiratory medicine 2007;101:1350-4.

5. Cosgrove GP, Bianchi P, Danese S, Lederer DJ. Barriers to timely diagnosis of interstitial lung disease in the real world: the INTENSITY survey. BMC pulmonary medicine 2018;18:1-9.

6. Hoyer N, Prior TS, Bendstrup E, Wilcke T, Shaker SB. Risk factors for diagnostic delay in idiopathic pulmonary fibrosis. Respiratory research 2019;20:1-9.

7. Pritchard D, Adegunsoye A, Lafond E, Pugashetti JV, DiGeronimo R, Boctor N, Sarma N, Pan I, Strek M, Kadoch M. Diagnostic test interpretation and referral delay in patients with interstitial lung disease. Respiratory research 2019;20:1-9.

8. Ryerson CJ, Vittinghoff E, Ley B, Lee JS, Mooney JJ, Jones KD, Elicker BM, Wolters PJ, Koth LL, King Jr TE. Predicting survival across chronic interstitial lung disease: the ILD-GAP model. Chest 2014;145:723-8.

9. Ley B, Ryerson CJ, Vittinghoff E, Ryu JH, Tomassetti S, Lee JS, Poletti V, Buccioli M, Elicker BM, Jones KD. A multidimensional index and staging system for idiopathic pulmonary fibrosis. Annals of internal medicine 2012;156:684-91.

10. Wells AU, Desai SR, Rubens MB, Goh NS, Cramer D, Nicholson AG, Colby TV, Du Bois RM, Hansell DM. Idiopathic pulmonary fibrosis: a composite physiologic index derived from disease extent observed by computed tomography. American journal of respiratory and critical care medicine 2003;167:962-9.

11. Raghu G, Remy-Jardin M, Myers J, Richeldi L, Ryerson C, Lederer D, Behr J, Cottin V, Danoff S, Morell F. European respiratory society, Japanese respiratory society, diagnosis of idiopathic pulmonary fibrosis. An official ATS/ERS/JRS/ALAT clinical practice guideline. Am J Respir Crit Care Med 2018;198:e44-e68.

12. Walsh SL, Hansell DM, editors. High-resolution CT of interstitial lung disease: a continuous evolution. Seminars in respiratory and critical care medicine; 2014: Thieme Medical Publishers.

13. Brenner DJ, Hall EJ. Computed tomography—an increasing source of radiation exposure. New England journal of medicine 2007;357:2277-84.

14. Picano E, Semelka R, Ravenel J, Matucci-Cerinic M. Rheumatological diseases and cancer: the hidden variable of radiation exposure. BMJ Publishing Group Ltd; 2014. p. 2065-8.

15. Fain S, Schiebler ML, McCormack DG, Parraga G. Imaging of lung function using hyperpolarized helium‐3 magnetic resonance imaging: review of current and emerging translational methods and applications. Journal of magnetic resonance imaging 2010;32:1398-408.

16. Mugler III JP, Altes TA. Hyperpolarized 129Xe MRI of the human lung. Journal of Magnetic Resonance Imaging 2013;37:313-31.

17. Mugler III JP, Altes TA, Ruset IC, Dregely IM, Mata JF, Miller GW, Ketel S, Ketel J, Hersman FW, Ruppert K. Simultaneous magnetic resonance imaging of ventilation distribution and gas uptake in the human lung using hyperpolarized xenon-129. Proceedings of the National Academy of Sciences 2010;107:21707-12.

18. Edelman RR, Hatabu H, Tadamura E, Li W, Prasad PV. Noninvasive assessment of regional ventilation in the human lung using oxygen–enhanced magnetic resonance imaging. Nature medicine 1996;2:1236-9.

19. Hatabu H, Tadamura E, Levin DL, Chen Q, Li W, Kim D, Prasad PV, Edelman RR. Quantitative assessment of pulmonary perfusion with dynamic contrast‐enhanced MRI. Magnetic Resonance in Medicine: An Official Journal of the International Society for Magnetic Resonance in Medicine 1999;42:1033-8.

20. Ley S, Ley-Zaporozhan J. Pulmonary perfusion imaging using MRI: clinical application. Insights into imaging 2012;3:61-71.

21. Grobner T. Gadolinium–a specific trigger for the development of nephrogenic fibrosing dermopathy and nephrogenic systemic fibrosis? Nephrology Dialysis Transplantation 2006;21:1104-8.

22. Glandorf J, Klimeš F, Voskrebenzev A, Gutberlet M, Behrendt L, Crisosto C, Wacker F, Ciet P, Wild JM, Vogel-Claussen J. Comparison of phase-resolved functional lung (PREFUL) MRI derived perfusion and ventilation parameters at 1.5 T and 3T in healthy volunteers. PloS one 2020;15:e0244638.

23. Bauman G, Puderbach M, Deimling M, Jellus V, Chefd'hotel C, Dinkel J, Hintze C, Kauczor HU, Schad LR. Non‐contrast‐enhanced perfusion and ventilation assessment of the human lung by means of Fourier decomposition in proton MRI. Magnetic Resonance in Medicine: An Official Journal of the International Society for Magnetic Resonance in Medicine 2009;62:656-64.

24. Fischer A, Weick S, Ritter CO, Beer M, Wirth C, Hebestreit H, Jakob PM, Hahn D, Bley T, Köstler H. SElf‐gated Non‐Contrast‐Enhanced FUnctional Lung imaging (SENCEFUL) using a quasi‐random fast low‐angle shot (FLASH) sequence and proton MRI. NMR in Biomedicine 2014;27:907-17.

25. Balasch A, Metze P, Stumpf K, Beer M, Büttner SM, Rottbauer W, Speidel T, Rasche V. 2D ultrashort echo‐time functional lung imaging. Journal of Magnetic Resonance Imaging 2020;52:1637-44.

26. Heidenreich JF, Weng AM, Metz C, Benkert T, Pfeuffer J, Hebestreit H, Bley TA, Köstler H, Veldhoen S. Three-dimensional ultrashort echo time MRI for functional lung imaging in cystic fibrosis. Radiology 2020;296:191-9.

27. Bauman G, Pusterla O, Bieri O. Ultra‐fast steady‐state free precession pulse sequence for Fourier decomposition pulmonary MRI. Magnetic resonance in medicine 2016;75:1647-53.

28. Yang B, Metze P, Balasch A, Stumpf K, Beer M, Rottbauer W, Rasche V. Reproducibility of functional lung parameters derived from free-breathing non-contrast-enhanced 2D ultrashort echo-time. Quantitative Imaging in Medicine and Surgery 2022;12:4720-33.

29. Klimeš F, Voskrebenzev A, Gutberlet M, Kern A, Behrendt L, Kaireit T, Czerner C, Renne J, Wacker F, Vogel‐Claussen J. Free‐breathing quantification of regional ventilation derived by phase‐resolved functional lung (PREFUL) MRI. NMR in Biomedicine 2019;32:e4088.

30. Wild JM, Marshall H, Bock M, Schad LR, Jakob PM, Puderbach M, Molinari F, Van Beek E, Biederer J. MRI of the lung (1/3): methods. Insights into imaging 2012;3:345-53.

31. Hansell DM, Bankier AA, MacMahon H, McLoud TC, Muller NL, Remy J. Fleischner Society: glossary of terms for thoracic imaging. Radiology 2008;246:697-722.

32. Chan TF, Vese LA. Active contours without edges. IEEE Transactions on image processing 2001;10:266-77.

33. Yu J, Xue Y, Song HK. Comparison of lung T2* during free-breathing at 1.5 T and 3.0 T with ultrashort echo time imaging. Magnetic resonance in medicine 2011;66:248-54.

34. Zapke M, Topf H-G, Zenker M, Kuth R, Deimling M, Kreisler P, Rauh M, Chefd'hotel C, Geiger B, Rupprecht T. Magnetic resonance lung function–a breakthrough for lung imaging and functional assessment? A phantom study and clinical trial. Respiratory research 2006;7:1-9.

35. Myronenko A. MIRT-Medical Image Registration Toolbox for Matlab. 2018.

36. Dournes G, Menut F, Macey J, Fayon M, Chateil J-F, Salel M, Corneloup O, Montaudon M, Berger P, Laurent F. Lung morphology assessment of cystic fibrosis using MRI with ultra-short echo time at submillimeter spatial resolution. European Radiology 2016;26:3811-20.

37. Gudbjartsson H, Patz S. The Rician distribution of noisy MRI data. Magnetic resonance in medicine 1995;34:910-4.

**Figures**

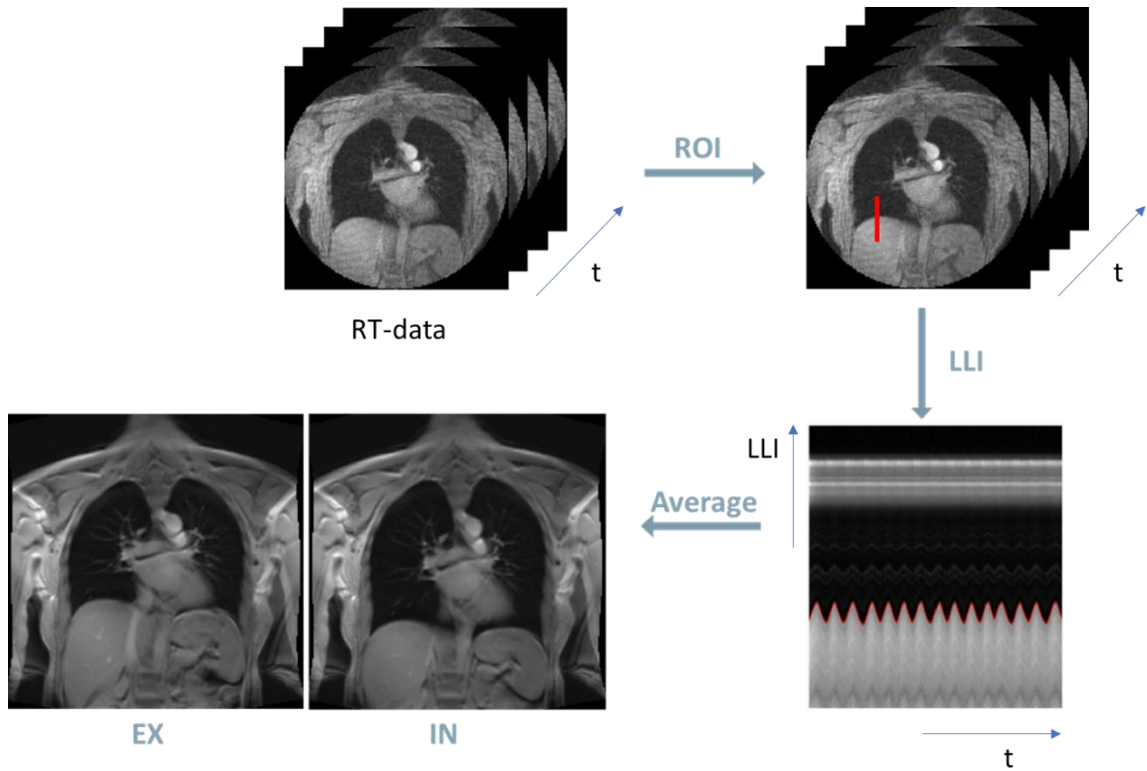

*Figure 1: A navigator-like signal for the respiratory motion was derived from the intensity profiles along the LLI in the real-time MRI data. Real-time images obtained in the same respiratory phase were averaged for SNR improvement. LLI: lung-liver interface; EX: Expiration; IN: inspiration.*

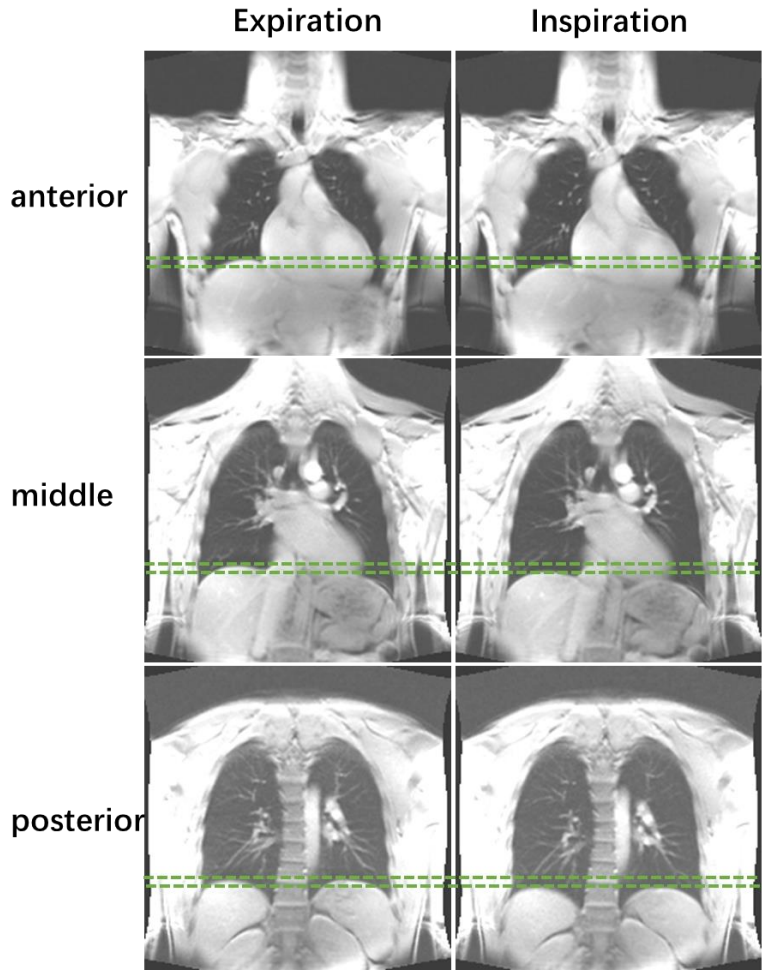

*Figure 2: Representative averaged real-time free-breathing MRI in expiration and inspiration in anterior, middle and posterior slices. Respiratory amplitudes were marked with green dashed lines.*

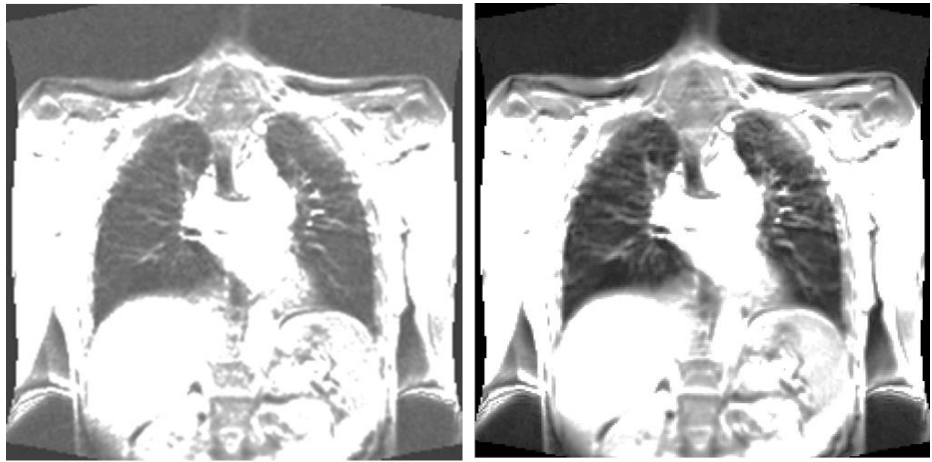
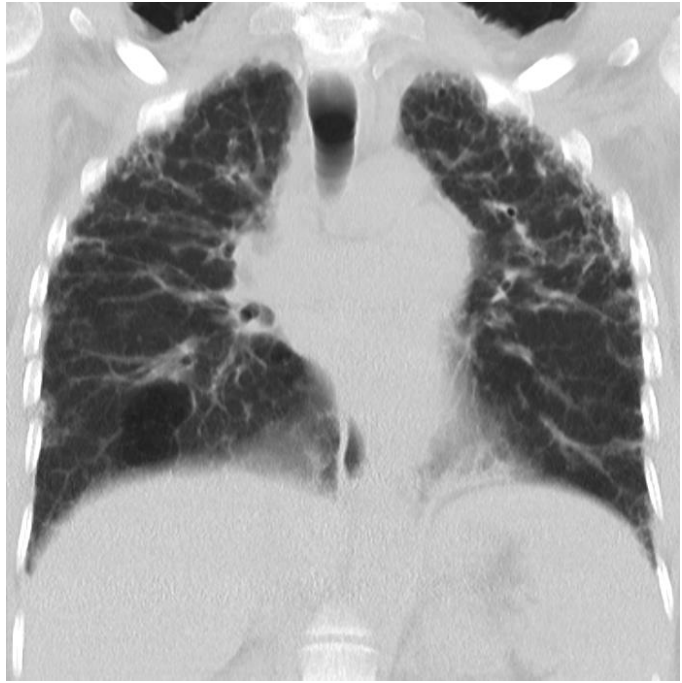

*Figure 3: Original a) and averaged b) (n=10) images of middle slice in inspiration of an ILD patient with idiopathic pulmonary fibrosis.*

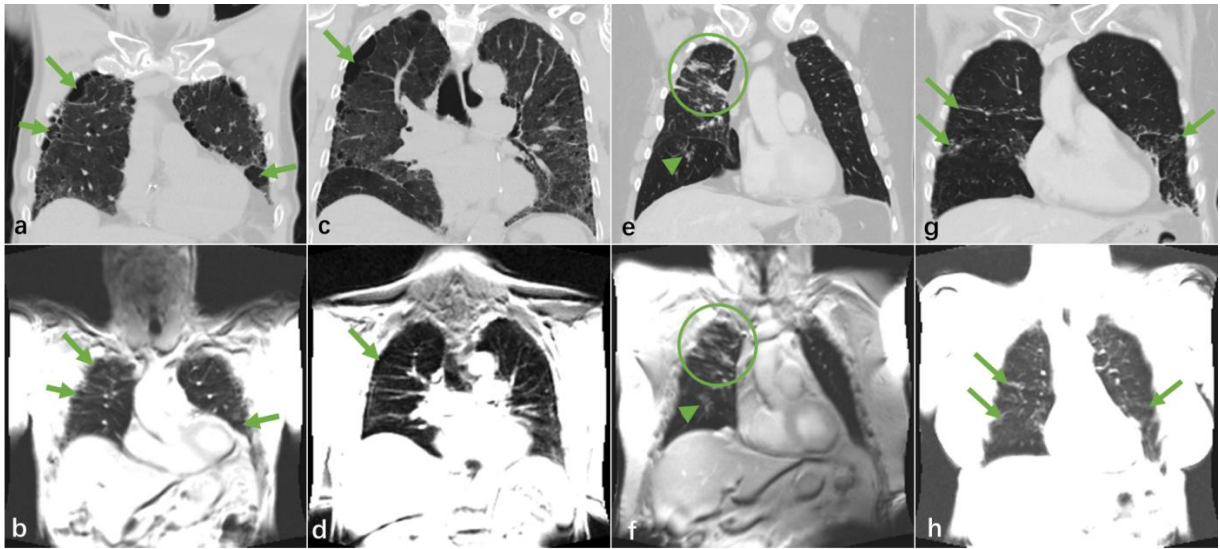

*Figure 4: CT (a,c), MRI (b,d) examinations in a 70-year-old male with idiopathic pulmonary fibrosis. Green arrows indicate areas of cyst and bullae. Diffuse ground glass opacities can be observed in both CT and MRI. CT (e) and MRI (f) examinations in a 65-year-old male with sarcoidosis. Green circles indicate areas of septal thickening and ground glass opacities. Green arrowheads marked nodules in both CT and MRI. CT (g) and MRI (h) in a 60-year-old female with the nonspecific interstitial disease. Green arrows indicate pericardial, lingual, and left upper lobe basal septal thickening.*

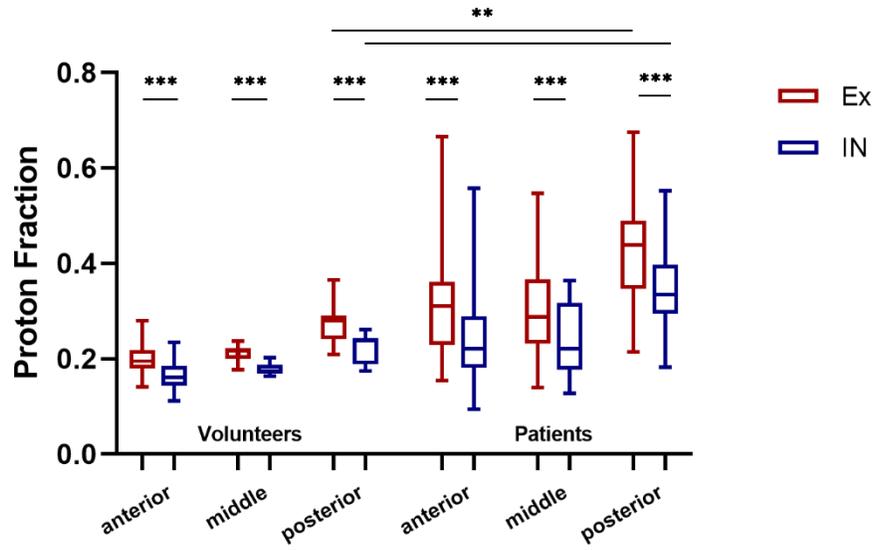

*Figure 5: Boxplots of proton fraction for expiration and inspiration from anterior to posterior of volunteers and patients with interstitial lung diseases. Significant differences (using an analysis of variance (ANOVA) test) are marked with an asterisk. Ex: Expiration; In: Inspiration; *(P < 0.05); **(P < 0.01; ***(P < 0.001).*

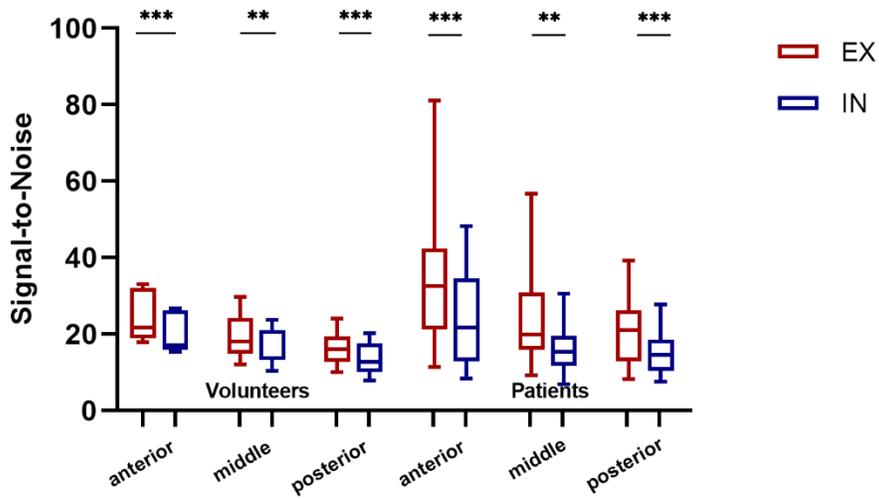

*Figure 6: Boxplots of signal-to-noise for expiration and inspiration from anterior to posterior of volunteers and patients with interstitial lung diseases. Significant comparisons (using an analysis of variance (ANOVA) test) are marked with an asterisk. Ex: Expiration; In: Inspiration; *(P < 0.05); **(P < 0.01).*

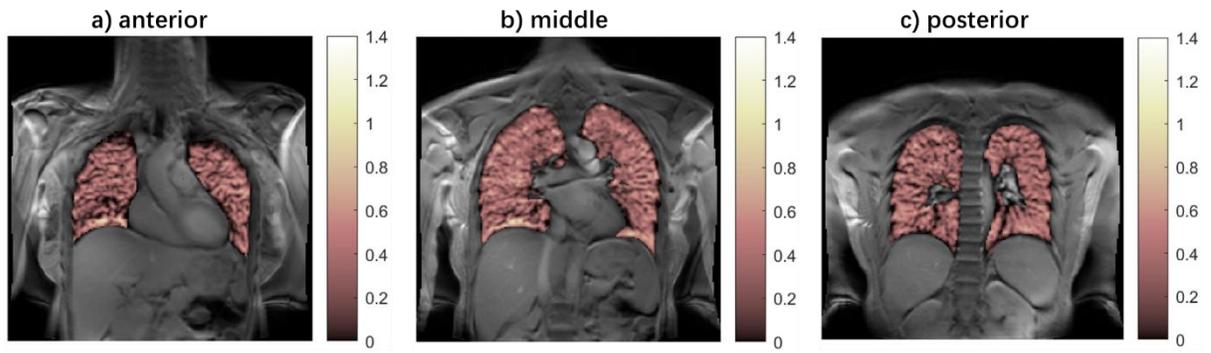

*Figure 7: Homogeneous fractional ventilation was observed in healthy volunteers. Anterior a), mid b) and posterior c) coronal slice*

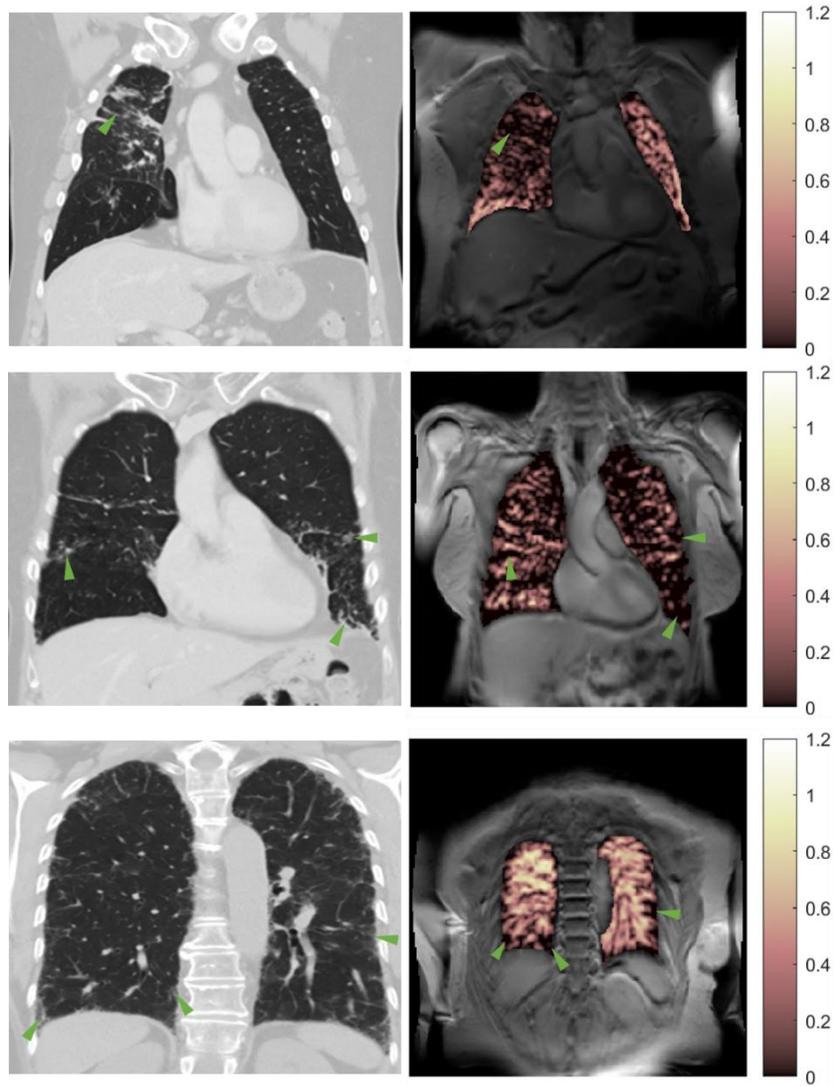

*Figure 8: Heterogeneous fractional ventilation was observed in patients with interstitial lung diseases. Regions with reduced fractional ventilation match areas with morphological changes. Clear reduction of FV can be detected in the area of fibrotic changes, ground glass opacities and pleural plaque (shown in green arrow head).*

**Tables**

**Table 1. Agreement between MRI and CT for all slices in patients with interstitial lung disease (blinded)**

|  | Kappa | P-value |
|---|---|---|
| Honeycombing | 0.333 | 0.074 |
| Septal lines | 0.590 | 0.018 |
| Reticulation | 0.636 | 0.006 |
| Consolidation | 0.846 | 0.001 |
| Ground glass opacity | 0.458 | 0.029 |
| Traction bronchiectasis | 0.333 | 0.074 |
| Nodule | 1 | <0.001 |
| Cyst | 1 | <0.001 |
| Pleural effusion | 1 | <0.001 |
| Emphysema | - | - |

**Table 2. Agreement between MRI and CT for all slices in patients with interstitial lung diseas (unblinded)**

|  | Kappa | P-value |
|---|---|---|
| Honeycombing | 0.600 | 0.009 |
| Septal lines | 0.818 | 0.001 |
| Reticulation | 0.750 | 0.002 |
| Consolidation | 0.846 | 0.001 |
| Ground glass opacity | 0.458 | 0.029 |
| Traction bronchiectasis | 0.818 | 0.001 |
| Nodule | 1 | <0.001 |
| Cyst | 1 | <0.001 |
| Pleural effusion | 1 | <0.001 |
| Emphysema | - | - |